\documentclass[runningheads]{llncs}
\usepackage{graphicx}

\usepackage{mathptmx,amsmath}

\usepackage{hyperref}

\begin{document}

\title{Modeling an Augmented Reality Game Environment to Enhance Behavior of ADHD Patients}
\titlerunning{Modeling an AR-Game for ADHD Patients}
\author{Saad Alqithami \and Musaad Alzahrani \and Abdulkareem Alzahrani \and Ahmed Mostafa}
\authorrunning{S. Alqithami, et. al.}
\institute{ Department of Computer Science, Albaha University, Saudi Arabia \\
\email{\{salqithami,malzahr,ao.alzahrani,amyosof\}@bu.edu.sa}}

\maketitle           

\begin{abstract}
The paper generically models an augmented reality game-based environment to project the gamification of an online cognitive behavioral therapist that performs instant measurements for patients with a predefined {Attention Deficit Hyperactivity Disorder} (ADHD). ADHD is one of the most common neurodevelopmental disorders in which patients have difficulties related to inattention, hyperactivity, and impulsivity. Those patients are in need for a psychological therapy; the use of cognitive behavioral therapy as a firmly-established treatment is to help in enhancing the way they think and behave. A major limitation in traditional cognitive behavioral therapies is that therapists may face difficulty to optimize patients' neuropsychological stimulus following a specified treatment plan, i.e., therapists struggle to draw clear images when stimulating patients' mind-set to a point where they should be. Other limitations recognized here include availability, accessibility and level-of-experience of the therapists. Therefore, the paper present a gamification model, we term as ``{\it AR-Therapist},'' in order to take advantages of augmented reality developments to engage patients in both real and virtual game-based environments. The model provides an on-time measurements of patients' progress throughout the treatment sessions which, in result, overcomes limitations observed in traditional cognitive behavioral therapies. 


\end{abstract}

\section{INTRODUCTION}
Attention Deficit-Hyperactivity Disorder is an increasing concern in the past few decades. It has undefined etiology as a heterogeneous developmental disorder leading to bias and extensive diagnostic evaluations when examining patients through traditional clinical interviews and rating of patients' behaviors~\cite{barkley1991ecological,abikoff1993teachers}. ADHD in underage patients can be observed in patients' hyperactivity and inability control their impulses and may have trouble paying attention which, in result, will intervene with their daily lives. In adulthood, patients with ADHD may have trouble managing time, being organized, setting goals, and holding down a job, which may lead to problems with relationships, self-esteem, and addiction. The treatment of many psychological disorders, such as ADHD, can be through a well-known type of psychotherapy called \textit{Cognitive Behavioral Therapy} (CBT). CBT involves patients in multiple psychosocial interventions in order to improve the status of their current mental health. This treatment requires patients to go through multiple sessions with specialized therapists. In ADHD, those sessions can be of increasing order of difficulty to help patients expand their cognitive capabilities to overcome current behavioral limitations.

ADHD has undefined etiology as a heterogeneous developmental disorder to involve hyperactivity and distractibility as well as difficulties with constant attention, impulsive control disorder and impaired cognitive flexibility, especially in problem solving and behavioral management~\cite{biederman2005attention}. Many studies have indicated the potential benefits of VR and AR exposure therapy for many types of mental disorders~\cite{parsons2007controlled,gorini2008second,beard2009survey,parsons2009virtual,bickmore2010response,meyerbroker2010virtual}. In a study by Parsons, et. al.~\cite{parsons2007controlled}, attention performance was compared between 10 children with ADHD and 10 normal control children in a VR classroom. The results showed that children with ADHD are more impacted by distraction in the VR classroom. In spite of that, Ben-Moussa, et. al. \cite{ben2017djinni} proposed a conceptual design of an exposure therapy system for patients with a social anxiety disorder. The proposd system integrates the AR and VR technologies through a simulated environment. It provides more effective exposure therapy solutions for patient with social anxiety disorder due to exploiting the benefits of AR and VR.

This highlights the importance of utilizing immersive technologies, e.g., Augmented Reality (AR) and Virtual Reality (VR), for their promising results that are stated in previous studies~\cite{parsons2007controlled,gorini2008second,beard2009survey,parsons2009virtual,bickmore2010response,meyerbroker2010virtual}. As the name suggests, AR-technology combines real and virtual contents, rendered in 3D, to be interactive in real-time \cite{azuma_survey_1997}. Whereas, VR-technology replaces the real world with a computer-generated graphics via head mounted display \cite{burdea_virtual_2003}. In other words, the user in VR environment totally isolated from the real-world while the AR optimizes the interactions in the real world \cite{billinghurst_survey_2015}. In AR, the environment is real but augmented with virtual objects from the system as it bridges the gap between real and virtual world in a seamless way~\cite{chang2010applications}. The information presented by the immersed virtual objects enhances a user's perception of- and interaction with- the real-world~\cite{azuma_survey_1997}. The rendering have much lower requirements in AR than in VR because VR completely replace the real world with the virtual environment which makes display devices used in AR have less stringent requirements than that used in VR systems. Also, the tracking and sensing requirements for AR are much stricter than those for VR systems~\cite{azuma_survey_1997}. This can be of great benefits for ADHD patients because it enables therapists to collect more information about the behavior of the patient which helps in the diagnosis and treatment procedures.

It is unfortunate that proposed solutions fail to overcome language and cultural barriers for diverse patients and to employ the power of augmented reality by rendering 3D objects and avatars rather than solid textual instructions leading to increase in patients' engagements and speeding-up the recovery. The believe is that there is no statistically significant difference between the ADHD patients who will be treated using this system and those who are treated by traditional CBT; although, online CBT may exceed traditional ones methods by accelerating recovery time and saving money and resources for both government and patients. This is due to achieving a concept of ``\textit{a therapist for each patient}" as the system mimics the therapist roles through an augmented reality development that provide it with features including: adaptiveness, smartness, responsiveness, and accuracy. Other advantages are availability, accessibility, and assurance of the therapist's level-of-experience which cannot be guaranteed in traditional CBT. 

Therefore, this paper models a system using a game pipeline and propose measurements that can be used to test the validity of using an online system as CBT, which we called ``\textit{AR-Therapist}''. The believe tackled here is that an increase in patient correct attention, e.g., when selecting a predefined object, contributes positively to their performance index which means they are following along with their treatment plan; the opposite should be true when they fail to achieve their assigned tasks. The result of this can be observed through applying the measurements we propose here on an augmented reality game.

\section{GENERIC DESIGN OF THE SYSTEM}
\begin{figure}[h]
  \centering
  \includegraphics[width=.65\linewidth]{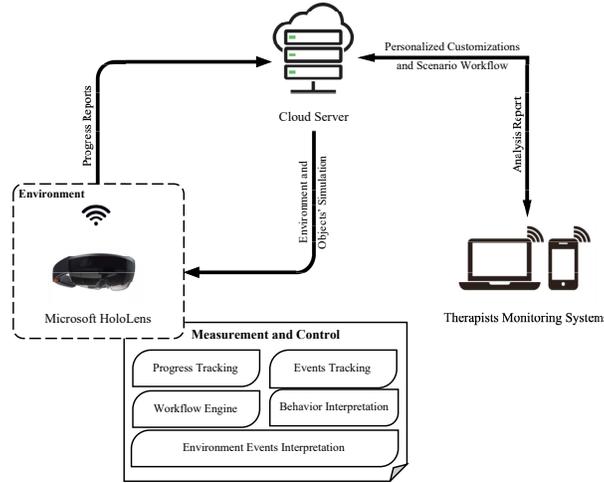}
  \caption{The System Architecture for AR-Therapist}
  \label{fig:arch}
\end{figure}

The high-level design, shown in Figure~\ref{fig:arch}, represents the main components of the AR-Therapist and the therapist-patient relationships within the online AR-environment.
 
Patients play-to-win the game that has been designed to attract their attention through the guidance of voice/sign instructions. Movement and actions are measured and then stored in a database to allow a therapist to monitor the treatment progress. Next, we present the logical flow presented in Figure~\ref{fig:arch}.

\subsection{Formal Game Pipeline:}
The AR-Therapist is based on the following 8-profiles for the generic assembly of the game pipeline.

\begin{enumerate} \setlength\itemsep{0em}
	\item ${\bf Treatment}$ is the whole treatment system (i.e., AR-Therapist). Its profile is a tuple of: $\langle$ Patient, Doctor, Game, \{Treatment-Program\} $\rangle$
		\begin{itemize}\setlength\itemsep{0em}
			\item {\it Patient}: Each patient will have his/her own profile. The profile has to be complete for the patient to join the treatment program. 
			\item {\it Doctor}: Each doctor will have his/her own profile.
			\item {\it Game}: The game has to be defined by the doctor considering patients current mental state and the disorder severity.
			\item {\it \{Treatment-Program\}}: Players will go through a treatment program that includes playing an AR-based game consisting of a set of levels.
		\end{itemize}

	\item ${\bf Patient}$: $\langle$ ID, Level, Performance-Index, \{Preference\} $\rangle$ 
		\begin{itemize} \setlength\itemsep{0em}
			\item {\it ID}: is a short identification as a name or a referral number used by the doctor to define a patient.
			\item {\it Level}: The level to where the patient has arrived in the treatment plan.
			\item {\it Performance-Index}: The current value of performance the patient has achieved throughout the game.
			\item {\it \{Preference\}}: The set of predefined preferences for a patient considering other psychological disorders that may affect current design and methodology of the treatment plan.
		\end{itemize}

	\item ${\bf Doctor}$: $\langle$ ID, Experience, Involvement $\rangle$
		\begin{itemize} \setlength\itemsep{0em}
			\item {\it ID}: The doctor has to have his/her own profile that is different from other therapists or psychological centers. This will give the doctor an access to the patient profiles and progress reports to allow for further evaluations.
			\item {\it Experience}: Experience level of the doctor is useful in allowing access to more complex/detailed data of the patients.
			\item {\it Involvement}: The level of engagement within the treatment process which allow the doctor to get involved in the game and in the reporting progress along the way of the patient assigned treatment.
		\end{itemize}

	\item ${\bf Game}$: $\langle$ Type, \{Level\} $\rangle$ 
		\begin{itemize} \setlength\itemsep{0em}
			\item {\it Type}: The type of the game to be played that has to be suitable for the patient. E.g., drag-and-drop and multiple-choices.
			\item {\it \{Level\}}: The game consists of a set of levels that have different levels of complexity.
		\end{itemize}

	\item ${\bf Level}$: $\langle$ \{Object\}, Max-Time, Effects $\rangle$ 
		\begin{itemize} \setlength\itemsep{0em}
			\item {\it \{Object\}}: Maximum set of objects used in this level.
			\item {\it Max-Time}: A predefined maximum time for the whole level to be completed or aport otherwise.
			\item {\it Effects}: Simple directional voice or instructions used for guidance in case of a remote following.
		\end{itemize}

	\item ${\bf Object}$: $\langle$ Shape, Size, Random-Location, Visibility $\rangle$ 
		\begin{itemize} \setlength\itemsep{0em}
			\item {\it Shape}: The structure of an object has to be predefined beforehand the start of a session.
			\item {\it Size}: The size of an object will depend on the location and closeness from the player focal point.
			\item {\it Random-Location}: The initial distribution of objects around the real-environment.
			\item {\it Visibility}: The appearance of one object after another.
		\end{itemize}

	\item ${\bf Treatment-Program}$: $\langle$ \{Game-session\}, Performance-Measures, Duration $\rangle$
		\begin{itemize} \setlength\itemsep{0em}
			\item {\it \{Game-session\}}: The set of game session to complete the treatment program.
			\item {\it Performance-Measures}: The performance in one session reports the correct, incorrect and uncompleted tried the patient has gone through in one session.
			\item {\it Duration}: The maximum treatment time for the whole treatment program. E.g., 20-min to complete the treatment.
		\end{itemize}

	\item  ${\bf Game-session}$: $\langle$ Level, Timer, Current-location, Number-of-tries $\rangle$
		\begin{itemize} \setlength\itemsep{0em}
			\item {\it Level}: The game level has to be defined beforehand. The initial level is defined in the patient profile and player can move from one level to another asynchronously depending on his/her achievement in the session and then the patient profile is updated. 
			\item {\it Timer}: To count the response time for the patients.
			\item {\it Current location}: To track current location of the patient for measuring closeness from objects within a session. 
			\item {\it Number of tries}: The repetition of tries within one session to include correct, incorrect and uncompleted tries. E.g., the number of collected target objects the patient has correctly collected in one session.
		\end{itemize}
\end{enumerate}

To this end, we have introduced Eight-profiles that when merged will best formulate the AR-Therapist model. Next, we show the process of combining those profiles into a conceptual model.

\subsection{General Conceptual Model:}
\begin{figure}[h]
		\centering
 		\includegraphics[width=.85\linewidth]{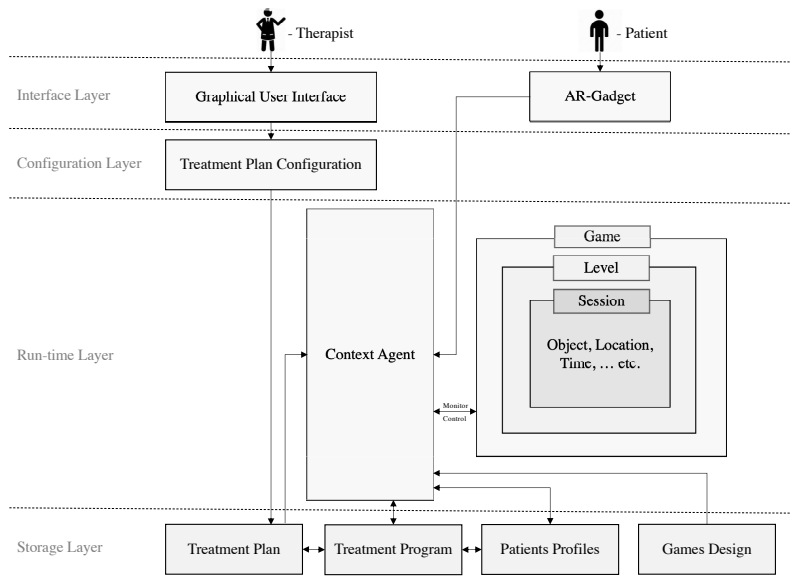}
		\caption{The general conceptual model for AR-Therapist}
		\label{fig:concept}
	\end{figure}

The conceptual model consists of Four layers: 1) interface layer, 2) configuration layer, 3) run-time layer, and 4) storage layer. Figure~\ref{fig:concept} depicts an architecture of the AR-Therapist based on the profiles presented in previous section.

\begin{enumerate} \setlength\itemsep{0em}
	\item \textbf{The Interface Layer} contains the interfaces used for accessing other layers. For instance, doctor can use a user-friendly interface to configure the treatment plan, whereas player interact with the game through Augmented Reality glasses. 

	\item \textbf{The Configuration Layer} consists of the ``treatment plan configuration'' component, where the doctor can add new treatment plan and configure the existence ones.

	\item \textbf{The Run-Time Layer} has the components that interact with the player while the game is running. These components are: the Context Agent component and the Game component.
		\begin{itemize} \setlength\itemsep{0em}
			\item[-] \textbf{The Context-Agent Component} retrieves the player treatment plan, and his/her performance in order to controlling the current game-session and guiding the player based on the treatment plan. Furthermore, the context agent capable of monitoring all the player's behaviors and interactions with the environment, logging ``Ethically'' the needed data, and calculating the player's performance following the measurements highlighted in the next section. Thus, the agent gains a deeper understanding of the patient to enrich the AR-Therapist model based on appropriate reasoning techniques. This in turns can be utilized in the future for suggesting the most optimal treatment plan for new patient (e.g. when facing a ``cold start issue''). In addition, logging the needed data and reasoning them will reveal some of hidden information that can be valuable for advancing research on treating patients with ADHD.

On the other hand, there is a persistent need for employing intelligent agent capable of performing the tasks explained above since the AR-Therapist model is used by the patients with the aid of their families and under the doctors' remote supervision (i.e. ambient assistant living ``AAL''). A prior use of AAL is observed in the literature for fall detection~\cite{chan_towards_2008}, and for monitoring elderly in their homes using video surveillance~\cite{foroughi_intelligent_2008}.

			\item[-] \textbf{The Game Component} involves levels of game; each level has its own games-sessions, and each game-session in turns contains the appropriate difficulty level that characterizes the maximum game-session time, used objects, and their locations.
\end{itemize}

	\item \textbf{The Storage Layer} involves components that store data about the treatment-plan, treatment-program, player profile, and game levels and objects. Each component follows an appropriate common specification for structuring the related data. Thus, the system's extensibility can be guaranteed as well as the integration with the existence E-Health systems.
\end{enumerate}

\section{Performance Measures:}
One of the most important issues in diagnosing and treatment of ADHD is to determine a set of accurate performance measures. These performance measures should have the ability to differentiate accurately between patients having ADHD symptoms and others who do not have it. According to psychiatric recommendations, it would be better to collect these performance measures from patients within different environments such as at homes or schools~\cite{morales2017psychometric}. There are many tests used to diagnose patients with ADHD~\cite{hall2016clinical}. Continuous Performance Tests (CPT) are the most popular laboratory-based test supporting clinical diagnosis~\cite{edwards2007estimates,vogt2011early,conners2000conners}. CPT is usually a computer-based test that aims to measure attention and impulsivity. CPT involves the individual and random presentation of a series of visual or auditory stimuli that changes rapidly over a period of time. Patients are informed to respond to the ``target'' stimulus and avoiding a ``non-target'' stimulus. The test provides summary statistics of performance parameters (e.g., response time, average response time, response time standard deviation, omission errors, and commission errors). These parameters have been shown to be useful in the detection of ADHD~\cite{rapport2000upgrading}.

An important limitation of traditional CPTs is low ecological validity~\cite{rodriguez2018comparison}. Ecological validity means the degree to which a performance test produces results similar to those produced in real life~\cite{negut2017virtual}. One approach to improve assessment methods which offers better ecological validity is CPTs based on VR, such as the Aula Nesplora test~\cite{climent2011aula}. Those approaches have an advantage of being more realistic and ecologically-valid environment while still having the ability to assess the degree of ADHD severity. Using AR instead of VR will further improve the ecological validity of the performed test.  Here, multiple performance measures is used to provide ADHD diagnostics and treatment assessment. Thus, let us assume that we have an AR-game that frequently present an interactive environment with the following assumptions:

\begin{itemize} \setlength\itemsep{0em}
	\item[] $T$: Total number of tries progressed in one session, which is a combination of:
		\begin{enumerate}
			\item[a.] Number of correct tries in one session, i.e., $C$,
			\item[b.] Number of incorrect tries (due to omission or commission errors) in one session, i.e., $I$, and
			\item[c.] Number of uncompleted tries in one session, i.e., $K$.
		\end{enumerate}
\end{itemize}

Then, the performance measures that will be used for providing ADHD diagnostics and treatment assessment include:
\begin{itemize} \setlength\itemsep{0em}
	\item {\bf Correct Response Times} ($CRT$): The percentage of measuring attention deficits for the time spent on the correct tries. CRT measures the time period that a patient takes to make a correct try (i.e., choose the target object in the try). The longer the CRT is, the more likely the patient has attention deficit and less time to focus on tasks. As a result, the patient takes longer time comparing to normal subjects when performing a task (i.e., choosing the target/correct object in our case). 
	\item {\bf Mean of the CRT} ($M$): To compare with correct response time to make sure they follow opposite relation to one another, i.e., $M = \sum_{i=1}^C {CRT_i}/{C} $.
	\item {\bf Standard deviation of the CRT} ($SD$): Indicative of impulsive and hyperactive symptoms of the patient. The higher the value of SD, the more probability that patient suffers from impulsivity/hyperactivity and difficulty in controlling his/her moves after a certain period of time. As a result, the patient starts periodically to move with no destination. I.e., $SD = \sqrt{\sum_{i=1}^C (CRT_i - M)^2/\{C-1\}}$. 
	\item {\bf Try time} ($\theta$): The maximum allowed time to complete one try within a session.
	\item {\bf Omission errors} ($OE$): The absence of any response during a try period to be used to measure inattention.
	\item {\bf Commission errors} ($CE$): The response to non-target stimuli which to be used to measure impulsivity.
	\item {\bf Engagement Factor} ($GF$): It indicates the patient engagement level with the game. In a case, a patient is considered to be engaged in the game if s/he continues to play the game. In contrast, the patient is considered to be not engaged if s/he stops the game before completing all tries in a session. Thus, $GF$ is defined as the number of correct and incorrect tries divided by the total number of tries in the session, i.e., $GF=  \{C+I\}/{T}$.
	\item {\bf Inattention Factor} ($IAF$): It represents the percentage of patient's inattention, i.e., $IAF=  {OE}/\{C+I\}$. In a case, the inattention of a patient increases when s/he makes Omission Errors ($OE$) in a session, i.e., when the patient does not choose any of the objects appearing to him/her. The number of uncompleted tries ($K$) in a session should be excluded when indicating $IAF$, i.e., $C + I = T - K$. 
	\item {\bf Impulsivity Factor} ($IMF$): Indicates the percentage of patient's impulsivity observed in his/her behavior within a session. $IMF$ is defined as the number commission errors divided by number of correct and incorrect tries, i.e., $IMF= {CE}/\{C+I\}  $. We also excluded $K$ when defining $IMF$. In our case, a patient who suffers from impulsivity will likely make more commission errors because impulsivity will prevent him/her from focusing when choosing an object. 
	\item {\bf Error Factor} ($EF$): It represents the percentage of the error rate during a session, i.e., $EF=\{OE+CE\}/\{C+I\}$. The error in the case includes omission and commission errors excluding $K$. Thus, $EF= IAF + IMF$.
	\item {\bf Correct Response Factor} ($CRF$): A percentage of total correct response-time relative to maximum allowed time for all correct tries, i.e., $CRF=  \{\sum_{i=1}^c CRT_i \}/\{C \times \theta\} $. In a case, $CRF$ should be negatively affected by the amount of time that the patient takes when s/he makes an incorrect try. Thus, we define CRF as the total summation of CRTs to the actual time of the game during the session ($GT$). In this case, $CRF$ will be $100\%$ if the patient makes all tries correctly. Otherwise, it will decrease depending on the total amount of time spent by the patient on $I$.

\item {\bf Performance Index} ($PI$): It reflects the single measure for the overall performance of the patient which depends on the correct response, error, and engagement factors.

	\begin{equation*}
		PI= \Bigg[ \frac{(1-CRF)+(1-EF)}{2} \Bigg] \times GF
	\end{equation*}
	
$PI$ is a composite score which measures the overall performance of a patient. In a case, the $PI$ is affected positively by the patients' $CRF$ and negatively by the patients' $EF$. In addition, we need to take into account different possible scenarios that can happen in the game session. One possible scenario is that the patient does not complete all the tries in the session. The patient can make one correct try and stop the game before finishing all the tries in the session. If we only considered the $CRF$ and $EF$, the $PI$ in the case would be the highest. In order to prevent this from happening, we consider the $GF$ in the definition of $PI$. Another possible scenario is that we have two patients who have the same $CRF$, $EF$, and $GF$ but different $GT$. In this case, they will have the same $PI$. However, the patient who has less $GT$ should have a higher $PI$. Thus, we considered the ratio of $GT$ to the maximum session time when defining the $PI$. 
\end{itemize}

\section{CONCLUSION AND FUTURE WORK}
The paper proposed a gamification system, called ``AR-Therapist'', as an online cognitive behavioral therapist to help in treating patients with a predefined ADHD symptoms. The purpose is to replace traditional CBT with a more advanced virtual one that may exceed traditional CBT methods by accelerating recovery time and saving money and resources. 
AR-Therapist achieves an excellent accessibility level to every patient in need, as it mimics the therapist roles through an augmented reality environment that provides it with features including: adaptiveness, smartness, responsiveness, and accuracy. Other advantages are availability and assurance of the therapist's level-of-experience which cannot be guaranteed in traditional CBT. We are working to implement a simulated augmented reality environment using a simple game, where the results should lead to an increase in patient correct attention (e.g., choosing a predefined object) which contributes positively to their performance index, i.e., they are following along with their treatment plan; the opposite is correct when they fail to achieve their assigned tasks. Future research should provide extensive empirical case studies that will support this area of research with a rich and valuable data when collected ethically and will introduce interesting findings when analyzed.

\section*{ACKNOWLEDGMENT}
This work was funded by the Deanship of Scientific Research at Albaha University, Saudi Arabia [grant number: 1439/4]. Any opinions, findings, and conclusions or recommendations expressed in this material are those of the author(s) and do not necessarily reflect the views of the Deanship of Scientific Research or Albaha University.

 \bibliographystyle{splncs04}
 \bibliography{Alqithami-B234}
\end{document}